\documentclass[review]{elsarticle}
\usepackage{amsmath}
\usepackage{graphicx}
\usepackage{hyperref}
\journal{Materials Chemistry and Physics}

\begin{document}

\begin{frontmatter}

\title{Transformation of a Metal-organic Framework for Tuned Catalytic Activity}

\author[ads]{Jiangtian Li$^{*,}$}
\cortext[cont]{Authors contributed equally}
\author[ads]{Terence Musho$^{*,**,}$}
\cortext[corr]{Corresponding Author: Email: tdmusho@mail.wvu.edu}
\cortext[cont]{Authors contributed equally}
\author[ads]{Joeseph Bright}
\author[ads]{Nianqiang Wu}
\fntext[ads]{Mechanical and Aerospace Engineering Department, West Virginia University}
\address{P.O. Box 6106, Morgantown, WV 26506-6106}

\begin{abstract}
Metal-organic frameworks (MOFs) are an attractive substrate for catalytic reactions due to the high area density of reaction sites and the ability to tailor an array of material attributes. This study focuses on a thermally stable crystalline UiO-66(Zr) MOF structure and the modulation of the electronic structure using two strategies to improve the catalytic conversion and selectivity of benzene alcohol to benzedehyate. Those two strategies include the functionalization of the organic struts with branched ligands and manually creating structural defects with unsaturated organic linkers. A combination of computational and experimental results provide evidence of improved catalytic activity of MOFs via these two approaches. Functional groups attached to the main organic strut modify the electronic environment of the photoactive aromatic carbon and thereby decrease the optical band gap by 1eV. Whereas the introduction of structural defects due to the organic linker desaturation provided a shift in the HUMO as a result of the decrease in strut coordination with the inorganic knots. \end{abstract}

\begin{keyword}
Metal-Organic Frameworks, Desaturation, Catalyst, Density Functional Theory
\end{keyword}

\end{frontmatter}
\newpage
%\linenumbers

\section*{Introduction}
The past two decades witness the rapid growth of a new class of porous materials, Metal-Organic Frameworks (MOFs), which are constructed with organic struts and inorganic knots in a highly periodic crystalline solid. Since its emergence, MOFs have attracted tremendous interest due to their structural diversity and flexible functionality in one single material, which endows MOFs potential applications in many areas such as gas capture/storage/separation, chemical sensing, and even drug delivery~\cite{1,2,3,4,5,6}.

Thanks to the high surface area and the permanent porosity, MOFs also build up an ideal platform for developing a new generation of heterogeneous catalysts.~\cite{6,7,8,9} The commonly employed approaches to control the catalytic activity in the porous materials like traditional porous zeolite is to engineer the surface functions and manipulate the structure defects that may bare variable functional active sites like Lewis Acid and/or Lewis Base. ~\cite{6,10} Over the traditional porous materials, MOFs exhibit the exceptional superiority on the chemical variability by introducing the molecular catalysts for desired catalytic reactions. These catalytically active molecules could be directly incorporated into MOFs during the synthesis process by being pre-linked on the organic linkers or post-grafted on the framework after the formation of MOFs. Theoretically, both approaches ensure the completely isolated and homogeneous dispersion of these molecular catalysts across the entire frameworks with the extreme accessibility for each functional moiety, which prevents the aggregation and the related deactivation, as well as the solubility concern that always disturbs the molecular homogeneous catalysis systems.~\cite{6,11,12} The versatility of molecular catalysts imparts the possibility to better control the selectivity and targeted catalytic reactions over MOF-based catalysts. 

Alternatively, exclusive the catalytic activity originating from the molecular moieties, some MOFs materials demonstrated superior catalytic activity from the metal ions at the inorganic nodes that could be made catalytically active by removal of the solvent ligands resulting in the coordinately unsaturated metal ion sites as the catalytic centers.~\cite{13,14,15} Analogous activity exactly happened to zeolite materials. Unfortunately, the understanding of the nature of such active sites in MOFs is at its early stage, and it still remains unclear how to sterically define and experimentally modify these sites, or how the sites’ activity can be further strengthened.~\cite{6} Remarkable insights are definitely required to interpret the intrinsic catalytic properties of the unsaturated metal nodes in MOFs. 

Another case in exploiting metal nodes’ catalytic activity in MOFs does not require the open metal ions, whereas treats MOFs as semiconductors that can be activated with light illumination. In such case, the organic linkers function as antennae and inject the photogenerated electrons to the metal nodes where catalytic reactions happen. Zn, Ti, and Zr-based MOFs have been proven photocatalytic active for organic dyes degradation, water splitting for H$_2$ generation and CO$_2$ reduction.~\cite{16,17,18,19} MOFs, as a new kind of photocatalyst, offer significant advantages over the conventional metal oxide photocatalysts: (i) high surface area and uniform pores facilitate the diffusion and adsorption of substrates; (ii) the redox center is highly exposed, leading to the increased reaction rates; and (iii) organic functionalization provides the flexibility for tailoring the electronic structure. However, most of these photoactive MOFs always fall in the ultraviolet light absorption range.~\cite{5,9} It remains a great challenge to harvest the solar spectrum as much as possible with stable MOF materials. 

With the prompt progress in this emerging field, MOFs with strong chemical and thermal stability have been successfully fabricated, which could definitely broaden their catalytic applications under much harsher conditions. As mentioned above, their catalytic activity stems from either the organic molecular catalyst or metal nodes. All these features enable MOFs as the combination of heterogeneous and homogeneous catalysts,~\cite{6,12} which makes MOFs very sensitive to the modifications of the organic linkers and inorganic knots that critically determines the final activity on the catalytic reactions. Herein we combine the theoretical prediction and experimental results to show these effects originating from the commonly used preparation processes, (i) organic linker modification by varying the side groups and (ii) unsaturated organic linker defects that are commonly observed with very small change in the preparation parameters.

\section*{Results and Discussion}
\begin{figure}
\includegraphics[width=1.0\columnwidth]{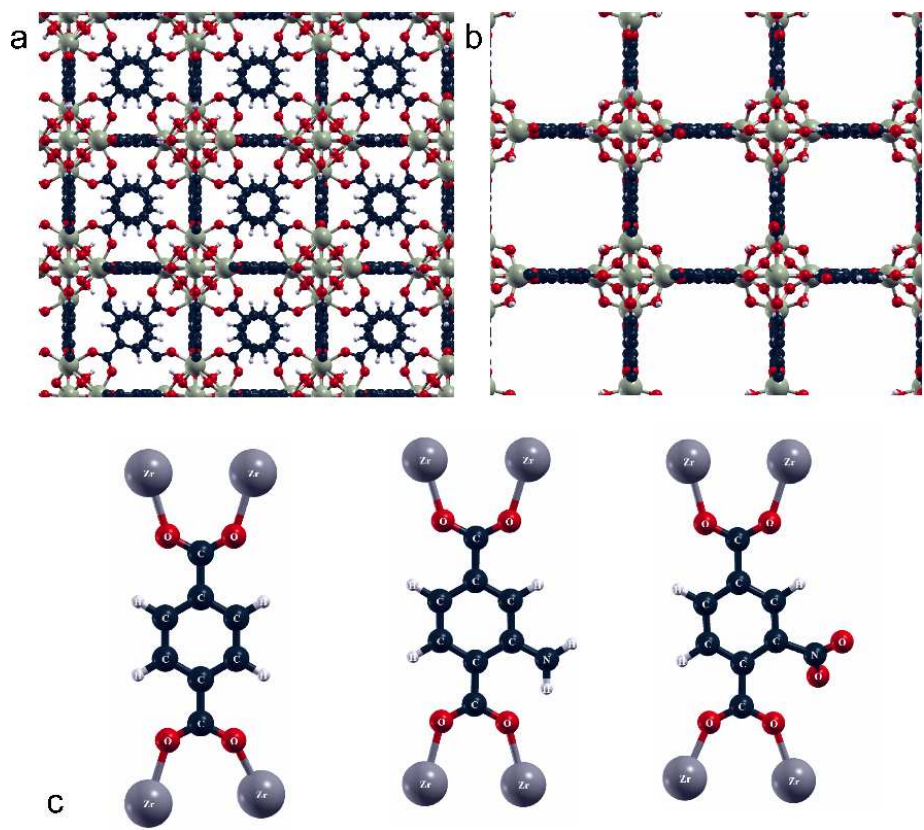}
\caption{Illustration of the face-centered-cubic crystal structure of UiO-66(Zr) MOF with inorganic knot Zr6O4(OH)$_4$ that will be bridged with organic strut 1,4-benzendicarboxylate (BDC). Subfigure (b) is the unsaturated MOF structure. Subfigure (c) illustrated the three different functionalizations that include BDC, BDC-NH$_2$, and BDC-NO$_2$. }
\label{fig:1}
\end{figure}

In this study, a Zr-based UiO-66 MOFs have been selected as the target MOF structure partially because of their exceptional chemical stability in strong acid aqueous solution, high thermal stability up to 300-400C, as well as the demonstrated catalytic activity as Lewis Acid substrate and photocatalytic activity for chemical fuel generation.~\cite{6, 16, 19,20,21,22} This conclusion could help guide the design for new MOFs materials. UiO-66(Zr) MOFs, with a nominal chemical formula [Zr6O4(OH)4][C6H4(COO)2]6, are built up with inorganic Zr6O4(OH)4 oxocluster knots bridged with 12 organic struts 1,4-benzendicarboxylate to form a face-centered cubic network (Figure~\ref{fig:1} a and b).\cite{23} The Zr6 inorganic unit is responsible for the robust UiO-66(Zr) chemical, mechanical and thermal stability, much more stable than most known MOFs.

\begin{figure}
\includegraphics[width=1.0\columnwidth]{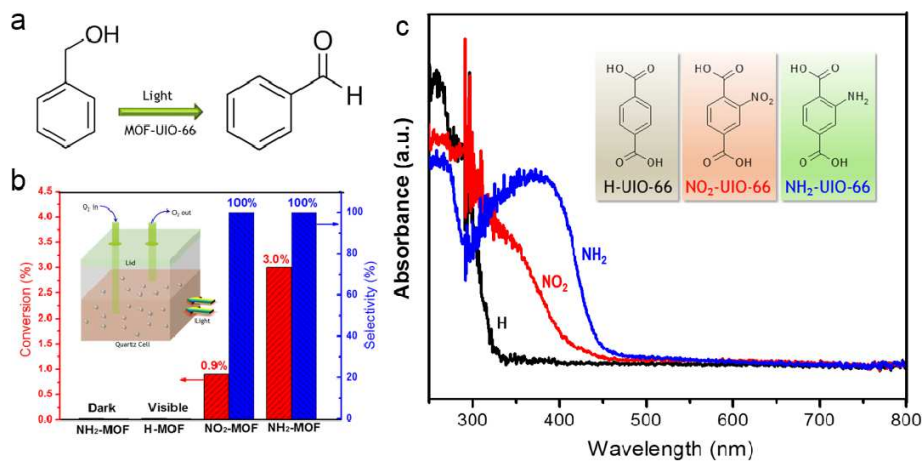}
\caption{(a) Organic transformation reactions studied in this paper. (b) Conversion efficiency and selectivity of benzene alcohol to benzedehyate for functionalized UiO-66 MOF materials under the visible light illumination.  The inset demonstrates the reaction cell for organic transformation. (c) depicts the UV-Vis light absorption of for UiO-66(Zr) with different branched groups on the aromatic ring of BDC.}
\label{fig:2}
\end{figure}

UiO-66(Zr) MOFs photocatalytic activity originates from the conjugated electronic structure of the organic linker leading to a UV band gap that generate moderately high potential photoexcited charge separation, which migrate towards the active sites.  Most photoactive MOFs have a band gap larger on the order of 3.4 eV or larger. Taking UiO-66(Zr) as an example, its band gap is estimated from UV-VIS light absorption spectrum is around 3.64eV corresponding to the absorption cutting edge at around 340nm (Figure~\ref{fig:2}c). The organic linkers can be lengthened to reduce the band gap energy, but the change mobility is limited. Also, these longer linkers rely on non-commercial organic linker precursors, which are difficult to synthesize.~\cite{24} Another approach for modulating the band gap is to exchange the metal cation. However, the stability and the redox levels limit the selection of metal cores for photoactive MOFs ~\cite{25}. A more favorable route to reduce the band gap and to modulate the electronic structure in MOFs is the substitution of the side groups onto the main organic linker. The conjugated $\pi$ electron structure in the organic linker is extremely sensitive to substitutions, which provides a simple yet effective modification procedure. ~\cite{26,27,28} Despite the pre-existing knowledge available from organic dyes, few studies have been conducted to investigate the effects of substituted linkers on the electronic band structure of MOFs. Few reports have given insight into the electronic band structure of MOFs. It remains a significant challenge to design and synthesize MOFs with tunable light capture capability for driving photocatalytic reactions. 

To realize the modulation upon the variation of the electronic structure, organic linkers with different branched groups, i.e. H, NO$_2$ and NH$_2$, corresponding to BDC, NO$_2$-BDC and NH$_2$-BDC, respectively, were used during the solvothermal synthesis process, which one-step yielded the UiO-66(Zr) MOFs materials with corresponding side functionalities, as shown in the inset in Figure~\ref{fig:1}c. The unchanged X-ray diffraction (XRD) patterns (Fig. S1) indicate that the substituted side groups did not alter the original face-centered-cubic crystal structure of UiO-66(Zr) MOFs. Indeed we observed the apparent color changes for these three UiO-66(Zr) MOFs. To quantitatively evaluate the absorption edge, we performed the UV-VIS absorption spectrum measurements, as shown in Figure~\ref{fig:2}c. As expected, the substitution of side groups results in the broadening of light absorption into the visible light region. H-UiO-66 has an absorption edge cutting at 340nm, corresponding to a band gap of 3.64eV in the deep UV region. NO$_2$ could significantly increase the light harvesting capability up to 425nm with a band gap of 2.92eV. NH$_2$ shows the maximum improvement in the light absorption to the 450nm corresponding to the band gap of 2.76eV. In light of the maintained crystal structure even with the protruded groups, the light absorption change must be related to the modulated electronic structures originating from the varying side groups. Table 1 is a summary of the experimental and DFT predicted band gaps for each of the structures. Both the experimental and DFT results convey the trend of decreasing band gap when the MOF linker coordination decreases. The reader should note that the DFT predictions are underpredicted due to the inherent assumption of DFT, however, the trends are similar.

\begin{table}
\begin{tabular}{ c c c c }
  Configuration & UV-Vis Cutting Edge (nm) & UV-Vis Band Gap (eV) &  DFT (eV)\\ \hline
  H-UiO-66 & 340 & 3.64 & 2.50 \\
  NO$_2$-UiO-66 & 425 & 2.92 & 2.10 \\
  NH$_2$-UiO-66 & 450 & 2.76 & 1.80 \\
\end{tabular}
\caption{Experimental and DFT prediction of the band gap for both the saturated and unsaturated MOF designs. The unsaturated case prove to decrease the band gap. The reader should note that the DFT band gaps are under-predicted due to the over analyticity of the functionals.}
\label{tab:1}
\end{table}

As mentioned above, the conjugated $\pi$ electron structure in the aromatic ring is extremely sensitive to substitutions, which subsequently affects the band gap structure of MOFs. To understand the influence of the bonding between the aromatic ring and the side functionalities, a ground state density functional theory (DFT) calculation was carried out to investigate the projected density of states and the origins of the modulation. Computational results validated that the band gap decreases in both NH$_2$- and NO$_2$-MOFs are prominently related to the sp2 hybridization between N atoms in functional groups (NO$_2$ and NH$_2$) and carbon atoms in aromatic rings. ~\cite{29,30} The different topology, however, determines the different ways that contribute to the electronic modulation. For NH$_2$ group, at the same time forming complementary π-bonds with the carbon atoms in the aromatic ring, the nitrogen atoms also form s-orbital hydrogen type bonding with the outlying pair of hydrogen on the functional group resulting in a mid-gap donor site, which contributes a valence state and results in a decrease in the band gap when compared to the non-functionalized BDC linker.~\cite{29} Whereas for NO$_2$ configuration, the un-fully satisfied nature of the N-O bond are ready to accept p-orbital electrons from the sp2 hybridized carbon atoms in aromatic ring, most likely functioning as an acceptor when comparing BDC-NO$_2$ to BDC, which leads to a slight modification of the p-orbitals, and then the combination contributes a valence state near the valence band maximum leading to a band gap decrease. In short, the band gap modulation was achieved and found to be highly influenced by the bonding nature of the functional group with the aromatic carbon ring.~\cite{29}

The model reaction for benzene alcohol to benzedehyate (Figure~\ref{fig:2}a) was used to test the catalytic activity of UiO-66 MOF materials under the visible light shining centered at 420nm. Typically 0.2g of UiO-66 powder was dispersed in 5mL of DMF with 200μL of substrate benzene alcohol. The catalytic reaction was performed under the light illumination with the bubbles of O$_2$. The reaction cell to conduct this experiment was shown in Fig. 2b. After 12 hours of reactions, the product was detected with Shimadzu GC-2014. First of all, as demonstrated in the previous report the sole product benzaldehyde was detected, ~\cite{19} indicating the hundred percent selectivity of UiO-MOF upon this reaction. As expected, NH$_2$-MOF and NO$_2$-MOF display the organic transformation efficiency of 3.0\% and 0.9\%, while H-MOF with negligible activity for this reaction. This result confirms that the catalytic activity for such organic transformation comes from the photocatalysis. Therefore, it is reasonable that NH$_2$ functionalized MOFs has maximum efficiency due to its strongest harvesting capability in this region, while the bald-BDC MOFs display zero activity because of the non-absorption for the visible light. The relative lower efficiency mainly due to the lower light power density of the light source. After the photon absorption with energy larger than the band gap of MOFs, the photogenerated charge carriers activate the molecular oxygens to create the reactive radicals that finally drive the organic transformation reactions.~\cite{19} The simple functionalization of the organic linkers gives rise to the modulation of the electronic structure and varying the band gap of MOF materials and subsequently the binding of the benzene alcohol.

The UiO-66 MOFs materials were prepared in a mild solvothermal process where organic linkers and inorganic knot precursors are dissolved in DMF and start self-assembly under solvothermal conditions. In this case, as talked above, each Zr6 oxoclusters should be fully coordinated and surrounded with 12 BDC-R organic linkers. ~\cite{23} Such UiO-66 MOFs materials as shown in Fig. 2 do not show catalytic activity on the benzedehyate reaction in dark, which indicates a perfect fully-coordinated UiO-66 MOFs crystals formed. However, when some extra coordinators like acetic acid and benzoic acid were added, there are systematically missed organic linkers for real materials. Such linker deficiency allows coordinative unsaturated sites on Zr to be identified as the superior active sites for Lewis Acid based reactions.~\cite{20,21,22,26} However, it remains a great challenge to precisely control the reaction activity, and even out of the control over the distribution of these active sites and acidity by varying the synthesis and post-treatment process.~\cite{6} It is not a surprise that MOFs were named as an opportunistic catalyst. Recently, a facile synthesis protocol to prepare UiO MOFs materials has been proposed by adding the concentrated HCl acid.~\cite{31} Different to other acid coordinators that would help slow down the reaction rate and enhance the crystallinity of MOFs materials, the presence of HCl speeds up the formation of UiO-66 materials more quickly by aiding in dissociating linkers from nodes. Through theoretical calculation and experimental results, it was assumed that four of 12 node linkers missing compared to the regular UiO-66 MOFs family materials, which would generate linker unsaturation defects and allow for more inorganic knots exposed, as compared in Figure 3.
 
In our this study, we also prepared this kind of unsaturated UiO-66-R MOFs materials by adding the concentrated HCl during the synthesis process. In the presence of HCl, we see the clear decolorization when compared to the regular UiO-66 MOFs materials. However, the face-centered-cubic crystal structure remained with decreased diffraction intensity. \cite{31}

To confirm the possible catalytic activity from the exposed Zr ions acting as Lewis Acid sites, we conducted the same organic transformation reaction (benzene alcohol to benzedehyate) with the same conditions except for no light illumination. As stated before, the regular UiO-66-R MOFs materials demonstrated no activity in driving this reaction in the dark (Fig. 4). Surprisingly, the unsaturated UiO-66-HCl demonstrated the unprecedented activity for this reaction with an efficiency up to 21\%. The product was detected to be pure benzedehyate, demonstrating the hundred percent selectivity. However, the unsaturated UiO-66-NH$_2$ and UiO-66-NO$_2$ exhibit almost same but very low activity for this reaction, which is completely different with the photocatalytic reactions in the regular UiO-66 MOFs materials. Without the photocarriers and the specific moieties that can drive the benzedehyate catalytic reaction, the superior activity in UiO-66-HCl should stem from the coordinatively unsaturated Zr ions serving as active Lewis Acid sites to activate this reaction. The dramatically suppressed activity with side NO$_2$ and NH$_2$ functionalities should be attributed to the electronic structure modulation that negatively affects the activity of these exposed Zr ions. In this process, molecular O$_2$ are required for generating active radicals to initiate the reaction.  So the interactions between O$_2$ and Zr ions in unsaturated UiO-66-R MOFs materials were calculated with DFT theory to elucidate the effect of the electronic modulation of the Lewis Acid sites’ activity. 

\begin{figure}
\includegraphics[width=1.0\columnwidth]{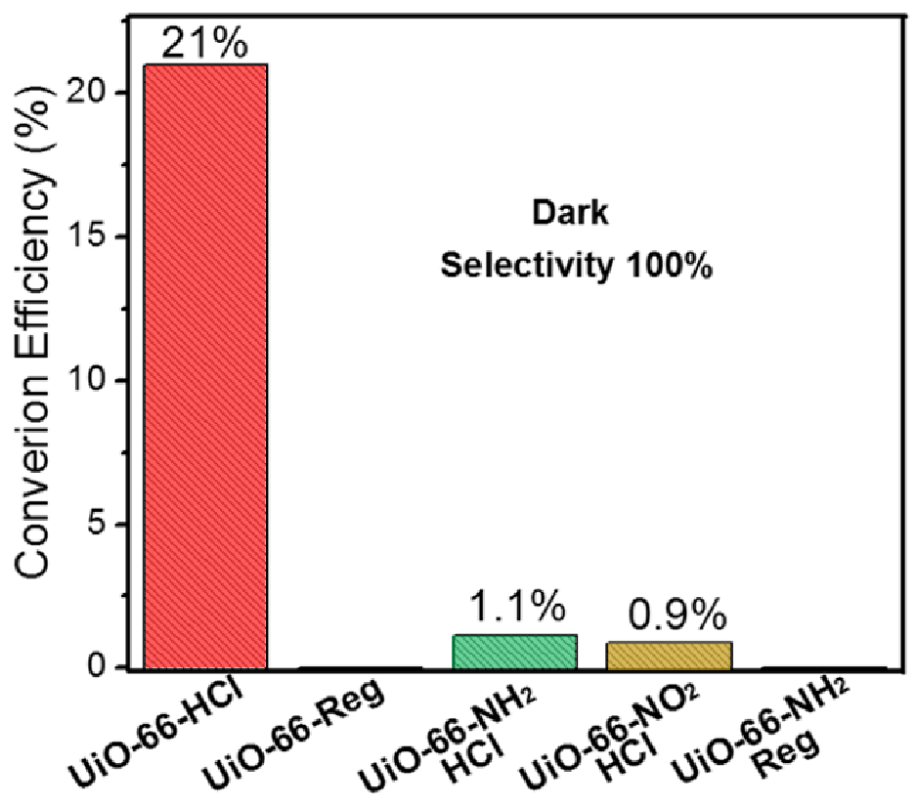}
\caption{Conversion efficiency and selectivity of benzene alcohol to benzedehyate for coordinately unsaturated UiO-66 MOF materials in light illumination. The dark selectivity was 100\%.}
\label{fig:3}
\end{figure}

\begin{figure}
\begin{center}
\includegraphics[width=0.7\columnwidth]{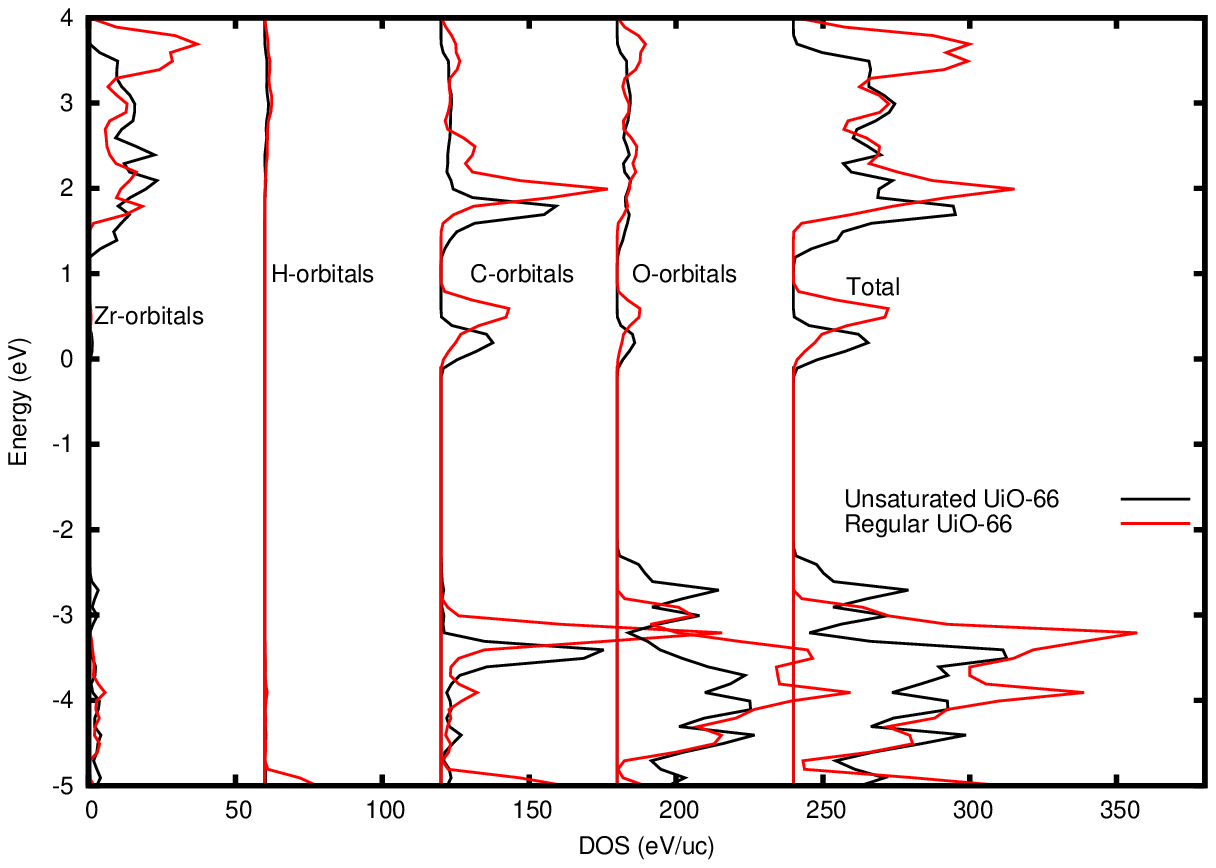}
\includegraphics[width=0.7\columnwidth]{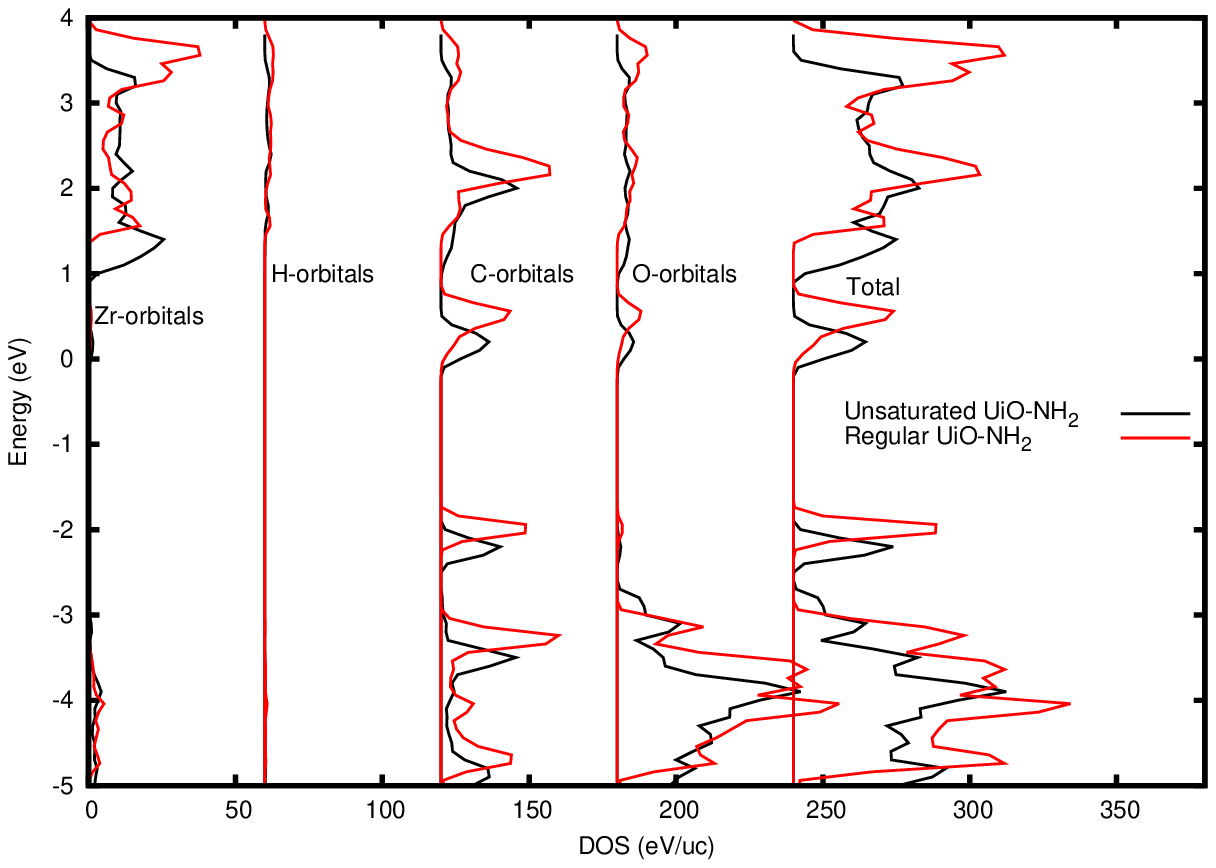}
\includegraphics[width=0.7\columnwidth]{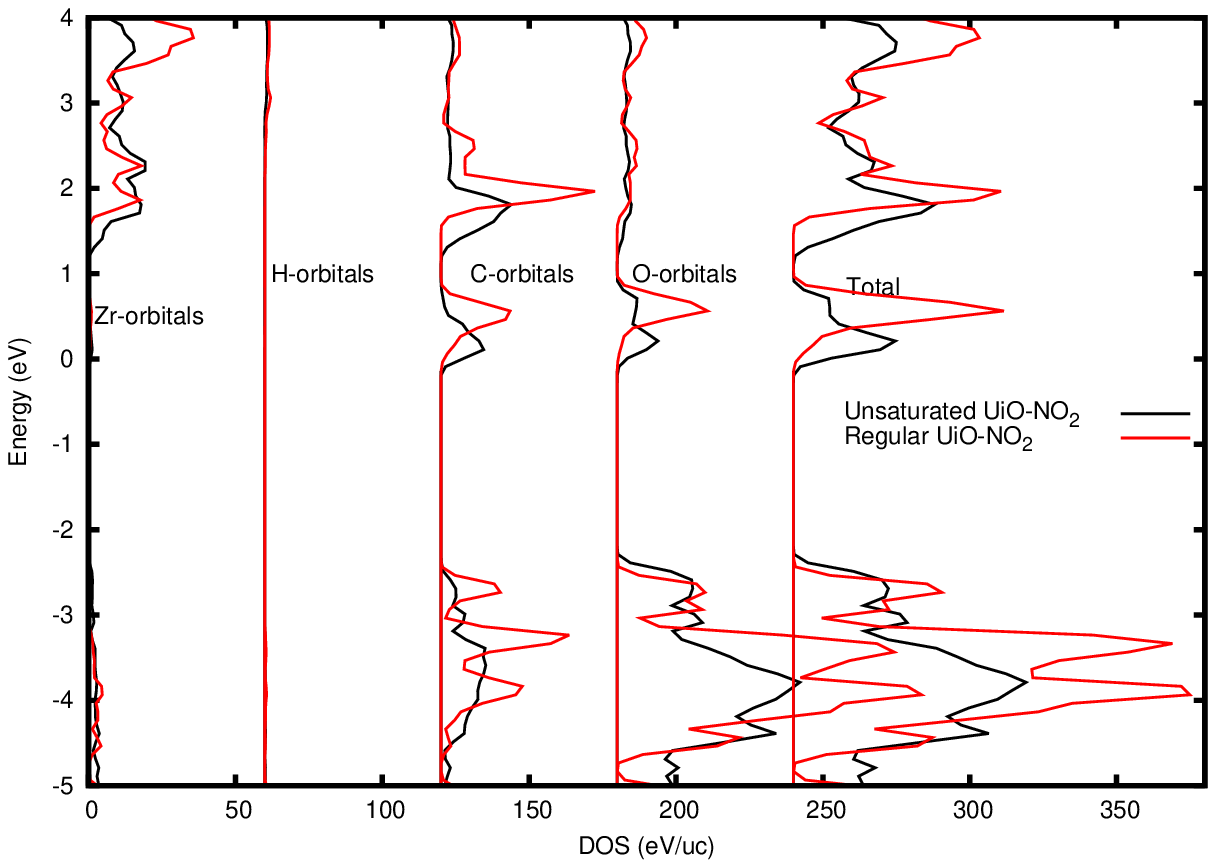}
\end{center}
\caption{Plots of the elemental projected density of states for both the unsaturated and saturated cases. Part a is the H-UiO(Zr)-66, b is the NH 2 , and c is the NO 2 . Unsaturating the structure results in a decrease coordination of oxygen and change in the valance states and ultimately a change in the Lewis Acid sites.}
\label{fig:4}
\end{figure}

To understand how the electronic structure is influenced by unsaturating the MOF structure DFT calculations were employed to determine the elemental projected density of states (PDOS). Figure~\ref{fig:4} is an illustration of the PDOS for all three cases in both unsaturated and saturated (regular) MOF designs. For all of the cases in Fig. 5 the Fermi energy is aligned to provide a relative comparison amongst all cases. All of the states below the Fermi energy are filled states and associated with the bonding of the MOF structure. All states above the Fermi energy are assumed unfilled or unoccupied. If we focus on the HOMO states below the Fermi energy and compare the density of states values for each species an understanding of the type of bonder can be ascertained. As the density of states is derived from the expectation value of the electronic wavefunctions, if there are complimentary peaks in the elemental projected density of states there is some certainty that these are covalent type states. This can be confirmed by the interaction and overlapping peaks in the density of states for both oxygen and carbon. This is a direct result of the classic sigma bonds that these two elements form the organic linker structure. More interesting is what unsaturating the structures does to the band structure. As is illustrated in Fig. 5 a and b there is a shifting of the oxygen density of states as the structure is unsaturated. Recall the oxygen atoms are situated at the intersection of the linker and the metalloid. In the unsaturated case the oxygen’s coordination is decrease. This shift in band structure is a result of a shift in the Fermi energy, which results in a change in the Lewis Acid. The decreased coordination of the oxygen results in the addition of a valance states influence the LUMO states. This can be further confirmed in all the cases of Fig. 5 by a change in the curvature of the LUMO levels from the introduction of several additional valence states. 

\section*{Conclusion}
MOFs combine the features for both heterogeneous and homogenous catalysis, which makes them are an intriguing platform to develop new catalysts for a wide range brand of potential applications. Due to the presence molecular catalytic moiety, the activity of MOFs materials are very sensitive to the electronic structure modulation even with very tiny structural variations, which offers a great tool to systematically manipulate the catalytic activity and selectivity of MOFs by coupling the other advantages such as diverse chemical availability on organic linkers and inorganic knots, as well as the tunable pore size and permanent porosity. We demonstrated theoretically and experimentally in this paper side group substitution on the main organic strut and coordinatively unsaturated metal ions acting active catalytic sites could be commonly used approach to modulate the electronic structure and finally alter the catalytic activity of MOFs materials.

\section*{Acknowledgements}
Acknowledgement is made to the Donors of the American Chemical Society Petroleum Research Fund (PRF\# 53490-ND10) for partial support of this research.
Acknowledgement is made to the Super Computing System (Spruce Knob) at WVU, which is funded in part by the National Science Foundation EPSCoR Research Infrastructure Improvement Cooperative Agreement \#1003907.

\section*{References}
\bibliography{reference}
\newpage
\section*{Supplemental Information}
\subsection*{Experimental}
The synthesis of UiO-66-R MOFs follows the literature.[S1,S2] Briefly, the same molar concentrations of ZrCl4 and the organic linker BDC-R were dissolved in DMF (1,4-benzenedicarboxylic acid for H, 2-amino-1,4-benzenedicarboxylic acid for NH2 and 2-nitro-1,4-benzenedicarboxylic acid for NO2, respectively). The resulting solution was transferred into a Teflon-lined autoclave for 48 hours at 120C. The precipitant was collected using a centrifuge and washed with DMF and methanol, and then re-dispersed in methanol for three days with gentle stirring. After that, the solid was collected and dried at 120C under vacuum.

The synthesis of these kinds of UiO-66 MOFs followed the procedure developed by Hupp group by adding concentrated HCl (35-38\%). Representatively, the proper amount of ZrCl4 was first dissolved in concentrated HCl aqueous solution, while the organic linker BDC-R (R=H, NO2, and NH2) were dissolved in DMF. Thereafter, two solutions mixed with ultrasonication to yield a clear solution. The resultant mixture was placed in an oven at 80C overnight. The precipitants were collected by centrifuge and washed with DMF twice and redispersed in methanol for one day. Finally, the powders were collected and dried under vacuum at 120oC overnight. It is supposed that there are at least three organic linkers lost for each unit cell. 
\begin{figure}
\begin{center}
\includegraphics[width=0.7\columnwidth]{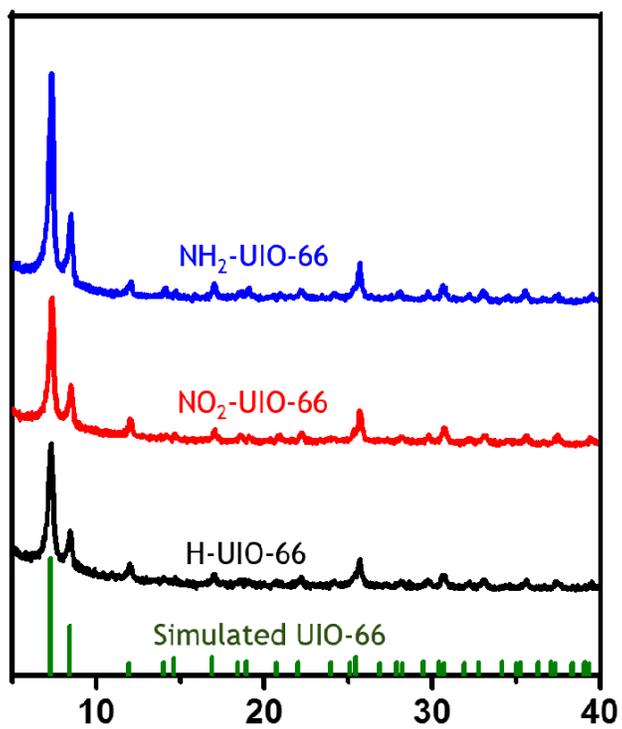}
\end{center}
\caption{XRD patterns for UiO (Zr)-66 MOFs with different side groups.}
\label{fig:s1}
\end{figure}
The UV-Vis absorption spectra of the as-prepared MOF materials were recorded on a Shimadzu 2550 UV-Vis spectrometer under the diffuse-reflection model using an integrating sphere (UV 2401/2, Shimadzu) coated with BaSO4.

The organic transformation catalysis testing was carried out in a flask containing 5mL of DMF, 200μL of benzene alcohol, and 100mg of UiO-66 MOF catalysts with stirring. All the reactions were conducted at room temperature with the continued O2 bubble as an oxygen source. The reaction solution was sampled with a regular time interval and then analyzed with Shimadzu 2014-GC equipped with a FID detector and an apolar CP-Sil 5CB column. For the photocatalytic activity, the reaction was conducted in a light box equipped with lamps with a visible light spectrum centered at 420nm. Whereas the other reactions conditions keep constant.

\subsection*{Computational}
Ground states were predicted by means of a density functional theory (DFT) approach [S4] following the approach described in previous studies [S2].  DFT calculations used functional form of the ultrasoft (US) pseudo-wave function that was based on Perdew-Burke-Ernzerhof (PBE) exchange-correlation function with a cut-off wave function energy of 816 eV (60 Ry), which provided most accurate and stable for the intended unit cell.  A Monkhorst-Pack k-point mesh sampling 4x4x4 grid with an offset of 1/2,1/2,1/2 was applied.  Van der Waals correction term [S5,S6] was incorporated to account for the Van der Waals interaction, which does include empiricism into the calculation. Scaling parameters were specified to be 0.7 and cut-off radius for the dispersion interaction was 900 angstrom. 

\subsection*{Supplemental References}
S1. J, Long, S. Wang, Z. Ding, S. Wang, Y. Zhou, L. Huang, X. Wang. Amine-functionalized zirconium metal-organic framework as efficient visible-light photocatalyst for aerobic organic transformations. Chem. Commun., 2012, 48, 11656-11658.\\
S2. T. Musho, J. Li, N. Wu. Band Gap modulation of functionalized metal-organic frameworks. Phys. Chem. Chem. Phys., 2014, 16, 23646-23653.\\
S3. M. J. Katz, Z. J. Brown, Y. J. Colon, P. W. Siu, K. A. Scheidt, R. Q. Snurr, J. T. Hupp, O. K. Farha. Chem. Commun., 2013, 49, 9449-9451.\\
S4. P. Giannozzi, S. Baroni, N. Bonini, M. Calandra, R. Car, C. Cavazzoni, D. Ceresoli, G. L. Chiarotti, M. Cococcioni, I. Dabo, A. D. Corso, S. de Gironcoli, S. Fabris, G. Fratesi, R. Gebauer, U. Gerstmann, C. Gougoussis, A. Kokalj, M. Lazzeri, L. Martin-Samos, N. Marzari, F. Mauri, R. Mazzarello, S. Paolini, A. Pasquarello, L. Paulatto, C. Sbraccia, S. Scandolo, G. Sclauzero, A. P. Seitsonen, A. Smogunov, P. Umari and R. M. Wentzcovitch, Journal of Physics: Condensed Matter, 2009, 21, 395502.\\
S5. S. Grimme, Journal of Computational Chemistry, 2006, 27, 1787–1799.\\
S6. V. Barone, M. Casarin, D. Forrer, M. Pavone, M. Sambi and A. Vittadini, Journal of Computational Chemistry, 2009, 30, 934–939.\\
\end{document}